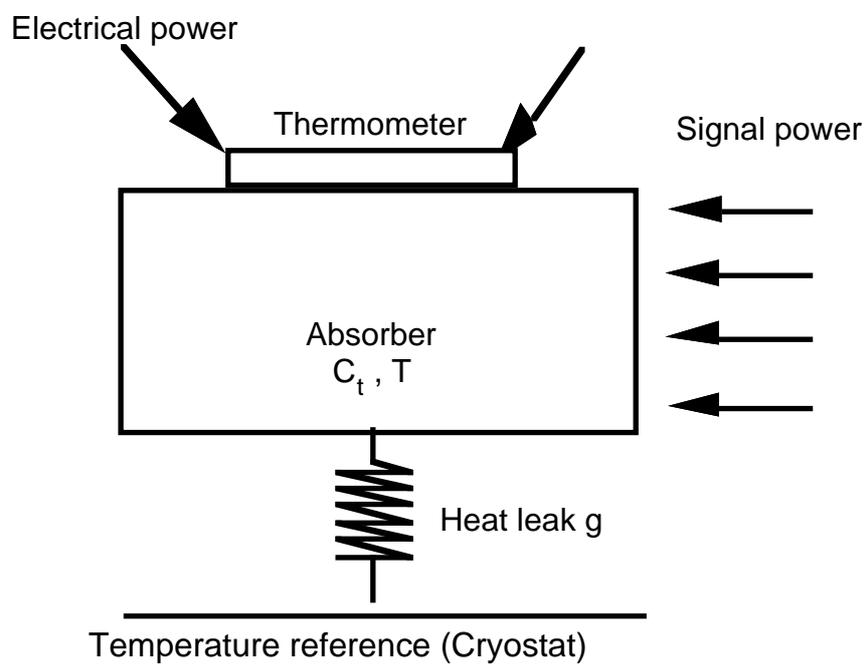

**Figure 1)**

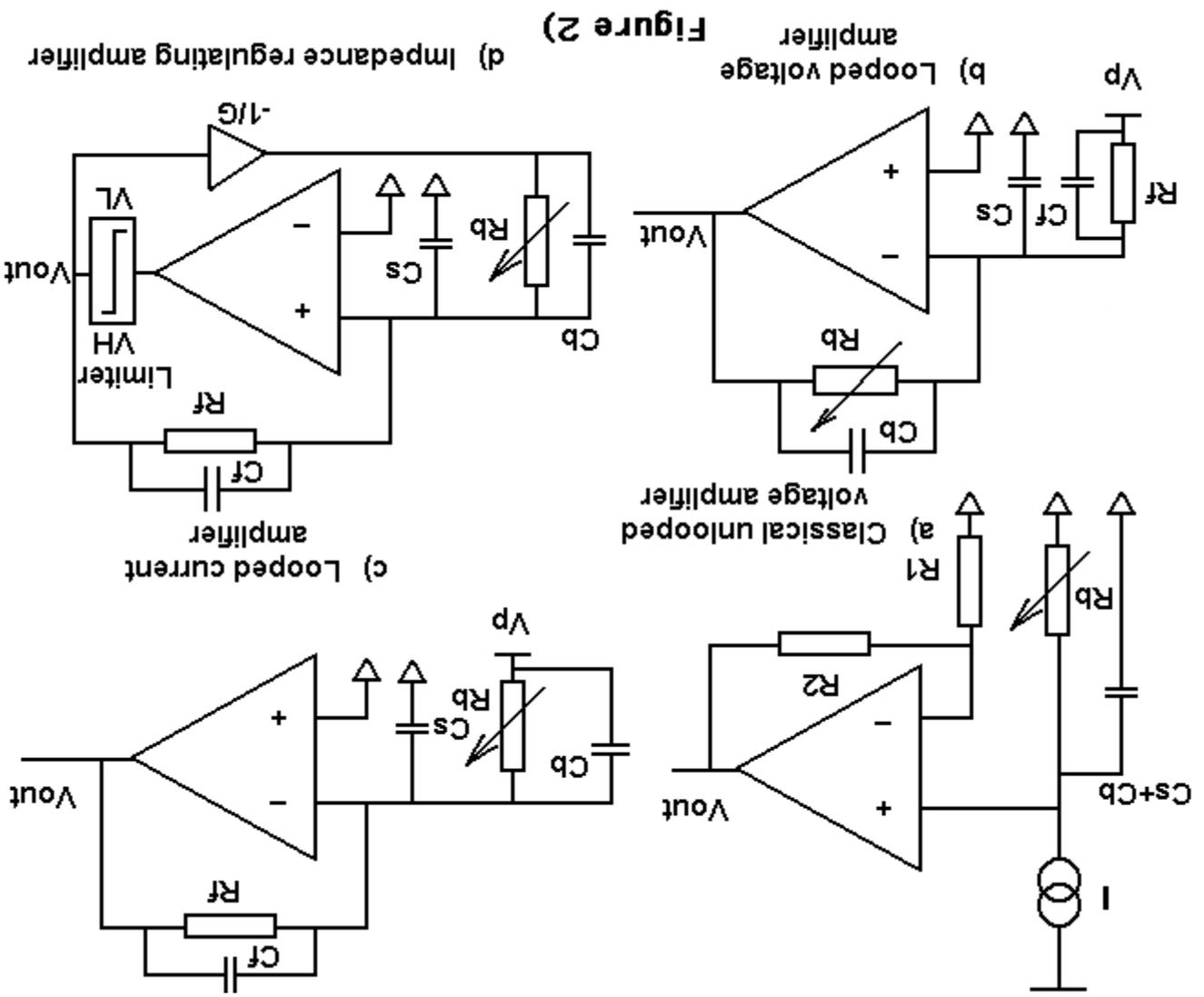

Figure 2)
a) Classical unlooped voltage amplifier
b) Looped voltage amplifier
c) Looped current amplifier
d) Impedance regulating amplifier

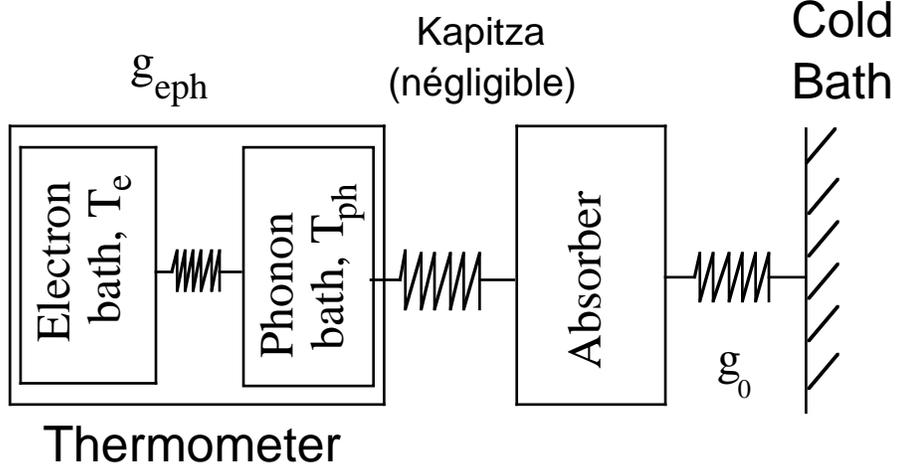

Figure 3)

# Low noise cryogenic electronics :
# preamplifier configurations with feedback on the bolometer

D. Yvon[1,*] and V. Sushkov[2,3]

[1]CEA, Centre d'Etude de Saclay, DAPNIA, Service de Physique des Particules, bat 141, F-91191 Gif sur Yvette Cedex.

[2]CEA, Centre d'Etude de Saclay, DAPNIA, Service d'Electronique et d'Informatique, bat 141, F-91191 Gif sur Yvette Cedex.

[3]Now at University of California Riverside, IGPP, Riverside CA 92521, USA.

*Corresponding author, Electronic mail address: yvon@hep.saclay.cea.fr, Tel: (33) 0169083911, Fax: (33) 0169086428

**Abstract:**

We have developed preamplifier configurations used to readout resistive bolometers such as those based on Neutron Transmuted Doped Germanium thermometers (so called NTD Ge thermometers)[1] or NbSi thin film thermometers [2, 3]. We introduce the impedance regulating preamplifier configuration. This configuration, is compared to previously proposed readout configurations with feedback on the bolometer [4]. It is shown that the impedance regulating preamplifier achieves Extreme Electrothermal Feedback without using transition edge thermometer and SQUID readout. Thus the detector and readout electronics are straightforwardly calibrated and can handle much larger event rates. Looped configurations simplify the design and improve performance of detection such as bandwidth. Finally, we analyse the impedance regulating amplifier noise contributions, with JFETs as front end device, relative to other configurations.

## I. Introduction:

Over the last ten years, low temperature detectors have undergone extensive development. They address detection problems that require high sensitivity, high-energy resolution and/or low thresholds such as X-ray and far infrared spectroscopy, double beta decay and Dark Matter searches. Their performance critically depends on the quality of the thermometer used, and on the readout electronics and wiring of the apparatus. This paper describes some of the electronics readouts we have developed for the EDELWEISS [5] and Planck Surveyor [6] collaborations.

We assume in the following that the thermometers used are resistive, of the order of 1 MΩ resistance, or more. The temperature increase of the bolometer induces a variation of the thermometer impedance. In this context, JFETs are used as front-end transistors to achieve low noise performance. We introduce a new preamplifier configuration: the impedance regulating preamplifier and discuss the properties of previously proposed preamplifier configurations with feedback on the bolometer [4] and of the classical unlooped voltage amplifier readout [7]. Preamplifiers have been developed or are under test for each of these configurations. First we will use a simple adiabatic model of bolometer with standard parameterisation to compute the bolometer's thermal behaviour. It is shown that feedback on bolometers yields significant changes in the detection behaviour such as shorter risetime and calibration of the detector. Thus, the impedance-regulating amplifier simplifies readout electronics, avoiding the complex additional circuits otherwise needed to achieve the same behaviour. The effect on signal to noise ratio is computed. We will summarise and compare the merits of the different readout configurations in the discussion. Detailed explanation of designs and performances of this electronic will be published later.

## II. Thermal behaviour of a bolometer: the adiabatic approximation.

Figure 1 shows a simplified schematic of the thermal structure of a bolometer, useful for computations. In this well know approximation [8], the bolometer is characterised by few parameters: its heat capacity $C_t$, its thermal conductivity to the cryostat: $g=dP_t/dT$ and the thermometer resistance dependence on temperature $R_b(T)$.

$P_t$ is the sum of $P_e$ from Joule effect of the current flowing through the resistive thermometer, $P_{ph}$ power of physical interest to be measured. We will use the dimensionless slope of the thermometer $\alpha$ defined as: $\alpha(T) = \dfrac{T}{R_b} \dfrac{dR_b}{dT}$

The thermal behaviour of the bolometer is described by the thermal differential equation:

$$C_t \frac{dT}{dt} = P_e - g_0(T-T_0) + P_{ph}, \qquad \text{with } g_0 = \frac{\Delta P_t}{\Delta T}, \ T_0 \text{ being the temperature of}$$

the cryostat cold plate.

In the limit of small temperature changes, we get:

$$C_t \frac{d\Delta T}{dt} = \Delta P_e - g\, \Delta T + \Delta P_{ph} \qquad \qquad \text{Equation (1).}$$

$$\text{with } g = \frac{d(\Delta P_t)}{d(\Delta T)}$$

The physical quantity computed from these parameters is the thermal relaxation time:

$\tau_{th} = C_t/g$  when $P_e=0$.

In an unbiased detector, when a particle interacts, releasing a power $E\,\delta(t)$ ($\delta(t)$ being Dirac's function) the change of the thermometer's resistance is expressed by:

$$\Delta R_b = \frac{R_b\, \alpha\, E}{T\, C_t}\, e^{-t/\tau_{th}}\, \theta(t) \qquad \qquad \text{Equation (2).}$$

For order of magnitude estimates, we will use the following typical values:

$C_t = 2\ 10^{-11}$ JK$^{-1}$, $P_e = P_t = 2\ 10^{-10}$ W, $T = 0.3$ K, $\alpha=-4$, $g = 3.3\ 10^{-9}$ WK$^{-1}$, $R_b=5\ 10^6\ \Omega$, $C_s= 100$pF, $R_f= 10^8\ \Omega$, where $C_s$ is the stray capacitance of the bolometer's wiring and the JFET. $R_f$ is the feedback resistor in the configurations that will be discussed later. Under these conditions $\tau_{th}= 6$ ms, $\omega_{th}=1/\tau_{th} = 165$ rad s$^{-1}$.

## III. Basic readout schematics and properties

Figure 2 shows the simplified schematics of the four preamplifiers configurations we will discuss. We will quickly review previously proposed configuration, then develop the impedance-regulating amplifier. In this paper, a looped configuration means that the bolometer's thermometer is part of the feedback loop of the preamplifier. We choose to discuss the interest of the impedance-regulating configuration by comparing it to the configurations previously proposed in literature. For the first order calculations performed in this paper, we consider that the operational amplifier is ideal, unless otherwise specified.

### A. Known configurations

#### 1. Unlooped Voltage amplifier:

In the first configuration Fig 2a), the thermometer is biased with a constant current source I. When a variation of $R_b$ occurs ($\Delta R_b$), it produces a voltage signal $\Delta V_b$ across $R_b$ given by: $\dfrac{dV_b}{dt} = \dfrac{1}{C_b + C_s}\dfrac{dq}{dt}$ where q is the charge in the capacitor $C_s$ and $C_b$. $C_b$ is the parasitic capacitor in parallel with the thermometer.

This signal is low-pass filtered by the $R_b(C_b+C_s)$ network and then applied to the input of the amplifier.

For an ideal operational amplifier, the response will follow variations of $V_b$ with amplification coefficient 1+(R2/R1). Solving this classical differential equation leads to:

$$\Delta V_b = I\,\Delta R_b \left(1 + \frac{R_2}{R_1}\right)\exp\left[\frac{-t}{R_b(C_s + C_b)}\right]$$

In the frequency domain we get:

$$\Delta V_{out} = \frac{I\Delta R_b}{1 + j\omega R_b(C_b + C_s)}\left[\frac{R_2}{R_1} + 1\right] \qquad \text{Equation. (3)}$$

Schematic of such a low noise preamplifier, as well as a discussion of the necessary compromises with the very low temperature cryogenic requirements are presented in reference [7].

## 2. Looped current amplifier

In the looped current amplifier (Fig 2c) the thermometer is biased with a constant voltage $V_p$. The ideal operational amplifier actively regulates the voltage at its inverted input to ground. $C_S$ does not charge. A variation of thermometer resistance $\Delta R_b$ induces a current variation $\Delta I = - V_p (\Delta R_b/R_b^2)$ that flows through $R_f C_f$ network, and thus in the frequency domain:

$$\Delta V_{out} = V_p \frac{\Delta R_b}{R_b^2} \frac{R_f}{1+j\omega R_f C_f} \qquad \text{Equation (4)}$$

This was first developed to handle high rate microcalorimeters [4] and then to study physics of ballistic phonons in large bolometers [2].

## 3. Looped voltage amplifier

The voltage amplifier (Fig 2b) is a variation of the current amplifier, where the position of the feedback resistor and the thermometer are inverted. The operational amplifier actively regulates the inverting input to ground. Thus the parasitic capacitor $C_S$ does not charge. The thermometer is biased with a constant current $I=V_p/R_f$, and we get in the frequency domain:

$$\Delta V_{out} = -\frac{V_p}{R_f} \frac{\Delta R_b}{(1+j\omega R_b C_b)} \qquad \text{Equation (5)}$$

## B. Impedance regulating amplifier

### 1. Introduction

The impedance-regulating amplifier Fig 2d) is more difficult to understand. Assuming the circuit is to be build around an ideal operational amplifier and neglecting stray and parasitic capacitors, the condition for the output not to be saturated is that the non-inverting input of the amplifier be kept to ground. Then

$$\frac{V_{out}}{R_f} = \frac{V_{out}}{GR_b} \qquad$$ where G is the ratio between $V_{out}$ and $V_b$ in the impedance

regulating amplifier configuration, immediately to the DC equilibrium criteria: $V_{out}$ grounded or

$$R_b = R_f / G.$$

The equilibrium output voltage is no longer computable from the electric circuit only. The system reaches equilibrium because the thermometer resistance changes when heated by an electric bias power. Let us assume, at startup, the thermometer is cold enough, so that $R_b \gg R_f/G$. When we power the preamplifier, it goes to saturation. The bolometer heats up, and the resistance of the thermometer drops, until the equilibrium criterion, $R_b = R_f/G$, is satisfied. Then the output voltage $V_{out}$ drops and stabilises to the equilibrium value $V_{eq}$ necessary to keep the temperature and resistance of the thermometer at their equilibrium values. We notice that if $V_{out} = V_{eq}$ is a stable state then, due to the central symmetry of I-V characteristic of bolometers, $-V_{eq}$ is another. In this readout configuration the preamplifier acts as a temperature regulator of the bolometer, and the feedback is *electrothermal*.

Let us define $P_{eq}$ the electric power dissipated into the thermometer at equilibrium $P_{eq} = \dfrac{V_{eq}^2}{G\,R_f}$. If the bolometer is heated, (by a variation of the incoming power, or a particle interaction), the output voltage will drop so as to keep the total power on the bolometer constant. Then in the small signal approximation:

$$\Delta P_{ph} = -\Delta P_e = -2\,\frac{V_{eq}}{GR_f}\Delta V_{out}. \qquad \text{Equation (6)}$$

We notice that measuring the output voltage of the preamplifier $V_{eq}$ and $\Delta V_{out}$, we measure, without the need of an external calibration, the physical power incoming on the bolometer (or energy of an event by integration), independently from the bolometer inner properties. The detector and readout electronics are naturally calibrated.

**2. Electronic response**

Let us now compute in more details the behaviour of the impedance-regulating configuration. In this paragraph we assume the preamplifier is composed of an operational amplifier of finite gain $A(j\omega)$, and input error voltage $\varepsilon(j\omega)$. Using conservation of charge in circuit 2d, and eliminating in the current flowing through $C_s$ in the equations we get :

$$V_{out} + Z_f I - \left[\frac{Z_f}{Z_s} + 1\right]\varepsilon = 0 \quad \text{and} \quad \varepsilon + Z_b I + \frac{V_{out}}{G} = 0$$

where $Z_s$ is the impedance of the capacitor $C_s$.

Eliminating I in these equations and using the relation $V_{out}(j\omega) = A(j\omega)\varepsilon(j\omega)$, we get:

$$Z_b = \frac{Z_f[1/G + 1/A]}{1 - \left[\frac{Z_f}{Z_s} + 1\right](1/A)} \quad \text{equation (7)}$$

where $Z_b$ and $Z_f$ are the impedances of the thermometer and feedback resistor.

In the limit of A large enough, this simplifies to $Z_b(j\omega) = Z_f(j\omega)/G(j\omega)$. To ensure this equilibrium condition at non-zero frequencies, $C_f$ will have to be adjusted to $C_f = \frac{R_b}{R_f} C_b$. The preamplifier will have to be tuned to a specific bolometer.

We now wish to compute the behaviour of the configuration when the thermometer resistance changes. Writing the three equations of charge conservation in figure 2d) and eliminating the current flowing through $C_s$ in two of them, we get:

$$I = -\left(\frac{V_{out}}{G} + \varepsilon\right)\frac{1}{Z_b}.$$

Injecting I in the third equation, we find: $Z_b = \frac{Z_f\left(\frac{V_{out}}{G} + \varepsilon\right)}{V_{out} - \left(\frac{Z_f}{Z_s} + 1\right)\varepsilon}$. Equation (6) shows that from an electric point of view $Z_b$ does not depend on $V_{out}$. We then differentiate the above equation with respect to $\varepsilon$, keeping $V_{out}$ constant. After Taylor expansion to first order, of numerator and denominator, we find:

$$dZ_b = \left\{Z_b\left[\frac{Z_f}{Z_s} + 1\right] + Z_f\right\}\frac{d\varepsilon}{V_{eq} - \left[\frac{Z_f}{Z_s} + 1\right]\varepsilon_{eq}} \quad \text{where } \varepsilon_{eq} \text{ is the value of } \varepsilon \text{ at equilibrium.}$$

Now using the relation $V_{out}(j\omega) = A(j\omega)\varepsilon(j\omega)$, and in the limit of A large we get:

$$dV_{out} = \frac{AV_{eq}}{\left[Z_b\left(\frac{Z_f}{Z_s} + 1\right) + Z_f\right]} dZ_b \quad \text{Equation (8)}$$

In the limit of $C_b = C_s = C_f = 0$, equation (8) simplifies to:

$$dV_{out} = \frac{AV_{eq}}{(G+1)R_b} dR_b \qquad \text{Equation (8bis)}$$

## IV. Dynamic performance

Let us now compute the behaviour (reaction and effective thermal bandwidth) of the different readout configurations. A heat signal is modelled by a power $P_{ph}=E\,\delta(t)$. This produces a decrease of the thermometer impedance according to equation (2). The observed relaxation time at the output of a biased bolometer does not match the thermal relaxation time of adiabatic model. This is due to the so-called electrothermal feedback effect.

We then solve equation (1), for the four different biasing conditions we are studying. We obtain: $\Delta T = \frac{E}{C_t} \exp\left(-\frac{t}{\tau_{eth}}\right)$, Equation (9) where $\tau_{eth}$ is the electrothermal relaxation time, which depends on the readout configuration. Let us define $H_{therm}(j\omega)$, $H_{ele}(j\omega)$ and $H_{tot}(j\omega)$ as

$H_{therm}(j\omega) = \Delta R_b/\Delta P_{ph}$, $H_{ele}(j\omega) = \Delta V_{out}/\Delta R_b$, and $H_{tot}(j\omega) = H_{therm}\, H_{ele}$. For all the readout configurations, we can write:

$$H_{therm}(j\omega) = \alpha \, \frac{R_b}{T_0 C_t} \frac{1}{s + 1/\tau_{eth}}$$

and $H_{ele}(j\omega)$ have been computed for all the configurations in the basics readout properties paragraph. From $H_{tot}(j\omega)$, we can compute the effective detection bandwidth of the readout configuration.

### A. Known configurations

#### 1. Unlooped voltage amplifier

Using an unlooped voltage configuration for a large thermometer impedance, the signal is low-passed, due to the parasitic capacitance of the readout connection and JFET input. $R_b = 10^7\,\Omega$ and $C_s$ of the order of 100 pF are typical of a readout connection in cryogenic environment. Then the time constant of electronic reaction is limited to:

$\theta = 2.2\ R_b\ (C_s+C_b) = 2.2$ ms.

Due to biasing with a constant current, $P_e = R_b\ I^2$. From equation (1) we get :

$$\tau_{eth} = \frac{C_t}{g - \frac{P_e \alpha}{T}}.$$ Metal to Insulator Transition thermometers have negative slope ($\alpha<0$), thus $\tau_{eth} = 3.3\ ms < \tau_{th}$. Biasing the thermometer shortens the bolometer relaxation time.

### 2. Looped voltage amplifier

From equation 5 we find that the signal is low-pass filtered with time constant $R_b\ C_b$. The electronic reaction time constant is $\theta = 2.2\ R_b C_b$. NTD-like thermometers have of the order of 20 pF parasitic capacitance. Thus the expected $\theta$ is 440µs, five time faster than the unlooped voltage readout configuration, but much slower than using the looped current amplifier configuration (see below), unless the parasitic capacitance of the bolometer can be made very small. Since the thermometer is biased through a constant current, we get :

$$\tau_{eth} = \frac{C_t}{g - \frac{P_e \alpha}{T}}, \qquad \text{and}\quad \tau_{eth} = 3.3\ ms < \tau_{th}$$

### 3. Looped current amplifier

From equation 4, we find that the signal is low-passed with the cutting angular frequency $\omega_f$. Resistors of high ohmic value, stable at a temperature of 20 mK with parasitic capacitance of 0.2 pF are available on the market. Using 10 MΩ resistors, the electronic reaction time constant of this detection configuration becomes 4.4 µs. However since the bolometer is biased through a constant voltage voltage, $P_e = \frac{V^2}{R_b}$. Computation of the electrothermal relaxation time gives:

$$\tau_{eth} = \frac{C}{g + \frac{P_e \alpha}{T}} \qquad \text{and}\quad \tau_{eth} = 32\ ms > \tau_{th}.$$

The higher is the electrical power $P_e$ delivered to the bolometer, the slower the thermal relaxation time is. Since $\alpha<0$, in this configuration $\tau_{eth}$ becomes infinite for $P_e = -gT/\alpha$, leading to thermal instability of the bolometer preamplifier chain.

## B. Impedance regulating amplifier.

### 1. Electronic reaction

From equation 8, assuming $C_s$ is much larger than $C_b$ and $C_f$ (experimental fact), we get: $dV_{out} = \dfrac{AV_{eq}}{R_b} \dfrac{\Delta R_b}{1+G+j\omega R_f C_s}$. This would imply a rise time of $R_f C_s/(1+G)$, but due to the large value of A, the preamplifier output enters saturation regime in a much shorter time scale and the thermal equilibrium is restored faster. Let us estimate this time constant by computing the time needed for the system to reach equilibrium when a step power function $\Delta P_{ph}$ enters the bolometer. The preamplifier stops rising when the input voltage reaches to the value: $\varepsilon = \dfrac{GR_f}{AV_{eq}} \Delta P_{ph}$. We integrate over time $\Delta I$ due to $\Delta P_{ph}$, assuming constant charging rate for $C_s$, and using the definition of $\alpha$. We get an order of magnitude of the corresponding time constant $\theta$:

$$\theta = \dfrac{R_f C_s}{A\alpha} \dfrac{T_0 g}{P_e} \quad \sim 1\ \mu s, \qquad \text{Equation (10)}$$

From this we conclude that time constant introduced by input stray capacitance is relatively short for this preamplifier configuration.

### 2. Electrothermal relaxation time

We now can compute the behaviour of the impedance-regulating amplifier to a physics event $E\ \delta(t)$, assuming no stray or parasitic capacitors. A large energy event would induce saturation of the preamplifier at zero output for a short time, proportional to its energy, until the thermometer cools enough and the output voltage becomes non zero:

$\Delta t_{sat} \sim \dfrac{E\ R_b}{V_b^2} \sim 0.8\ \mu s$, assuming E=1 keV event, and $P_e$=2 $10^{-10}$ W.

Then the thermometer impedance is very close to equilibrium value.

$V_b = -\dfrac{V_{eq}}{G} - \varepsilon = -\dfrac{V_{eq}}{G} - \dfrac{V_{eq}}{A}$. Thus $P_e = \dfrac{V_{eq}^2(1/G + 1/A)^2}{R_b}$.

Differentiating $P_e$ with $R_b$ and $V_{eq}$, and using equation (8bis) to replace $dV_{out}/V_{eq}$ in equation, we get: $dP_e = P_e \left(\dfrac{2A}{G+1} - 1\right)\dfrac{\Delta R_b}{R_b}$.

Using the definition of α this equation transforms to $dP_e = P_e \left(\dfrac{2A}{G+1} - 1\right)\alpha \dfrac{\Delta T}{T}$,

Replacing $P_e$ in equation 1 leads to $\tau_{eth} = \dfrac{C_t}{g + P_e \dfrac{\alpha}{T}\left(\dfrac{2A}{G+1} - 1\right)}$

If we define $\beta = \dfrac{P_e}{gT}$ (the order of unity) and we assume $A/(G+1)$ large compared to 1,

we finally get: $\tau_{eth} = \dfrac{G+1}{2A\alpha\beta} \tau_{th}$

$= 0.8\ \mu s$, assuming $A \sim 10^5$, $G=20$, and the numerical values above.

This shows that when using an impedance-regulating amplifier, bolometers become relatively fast detectors, able to handle counting rates of the order of magnitude of semiconductor diodes. Another readout system aiming at this property, using an unlooped voltage preamplifier and a long AC coupled feedback chain has been recently published [9].

However, the adiabatic thermal model of bolometer is incomplete. At very low temperatures (less than 200 mK), the electrons of the thermometer are only weakly linked to the phonons [10]. Figure 3 shows a more realistic thermal model of a bolometer at low temperatures. The electrons of the thermometer are weakly thermalised by phonons of the crystal lattice. This is quantified by an effective heat conductivity $g_{eph}$. The thermometer impedance is a function of its electron temperature. The impedance-regulating amplifier effectively regulates the temperature of the thermometer's electrons. When an event happens in the absorber, the maximum power flux induced on the thermometer is

$P_{max} = \dfrac{E}{g_{eph}C_{ph}}$, ($C_{ph}$ is the heat capacity of the absorber). The relaxation time becomes

$C_{ph}/g_{eph}$ (where $C_{ph}$ is the heat capacity of the absorber and we define $C_e$ as the thermometer heat capacity dominated by the electrons), preventing saturation of the preamplifier and slowing down the relaxation time. On the other hand, to be able to achieve microseconds relaxation time the bolometer will have to be designed to maximise $g_{eph}$. This can be achieved, as described in [3].

## C. Validation using PSPICE simulator

To demonstrate the validity and the limitations of the above computations and better understand the behaviour of the impedance-regulating preamplifier, we implemented a computer simulation using Pspice simulator. We simulated the dynamical non-linear thermal behaviour of the bolometer, according to the 'realistic' thermal model plotted at figure 3 and experimental data on NbSi thin film thermometer [2,3]. In the numerical example, we assume the bolometer is made of a sapphire wafer 7x12x0.6 mm$^3$. The NbSi thermometer is a 5x5mm$^2$ square, 10-nanometer thick, thin film whose resistance versus temperature law and electron-phonon coupling parameterisations have been fitted on data [11]. We used:

$$R(T) = R_0 \exp\left[\left(\frac{delta}{T_e}\right)^{0,64}\right] \qquad \text{Equation (11)}$$

and $P_{eph} = g_e V_s (T_e^{5.7} - T_{ph}^{5.7}) \sim g_{eph} \Delta T$ in the limit of small temperature differences, where $R_0 = 1900 \, \Omega$, delta = 7.72 Kelvin. $P_{eph}$ is the power flowing from the electron to the phonon bath, expressed in Watt. $V_s$ is the thermometer volume in cm$^3$ and $g_e = 280$ W cm$^{-3}$ K$^{-5.7}$ is a constant of the NbSi material used. $T_e$ is the thermometer electron temperature and $T_{ph}$ is the absorber phonon temperature. We assumed the heat leak between the Sapphire plate and the reference temperature was Kapitza thermal resistance dominated [12]:

$$P_K = g_K(T_{ph}^4 - T_0^4) \sim g_0(T_{ph} - T_0) \text{ in the limit of small signals, where}$$

$P_K$ is the power flowing through the Kapitza resistance in watt, $g_K = 10^{-8}$ W K$^{-4}$ and $T_0$ is the reference temperature of the refrigerator. Heat capacity of the electron bath and phonon bath where computed according to the formulae:

$$C_e(T_e) = c_e\, V_s\, T_e\ (J\ K^{-1}),\ \text{and}\ C_{ph}(T_{ph}) = c_{ph}\, T^3\ (JK^{-1})$$

where $c_e = 6\ 10^{-5}\ J\ K^{-2} m^{-3}$ and $c_{ph} = 10^{-9}\ J\ K^{-4}$.

Finally, we modelled the preamplifier as an operational amplifier of input offset $V_{os}$, an open loop gain $A_0$ of $10^4$, and a single pole with Gain Bandwidth Product of 2 GHz and we used G= 10. We used a limiter to prevent the output of the preamplifier from dropping to zero voltage, which happened to be an uninvited stable state of this preamplifier

### 1. Simulation output

Figure 4 shows the Pspice schematic we used to model the thermal behaviour of the bolometer and the electronics. Figure 5 displays the output voltage, the electron and phonon bath temperature as a function of time as calculated by Pspice. The thermometer resistance is directly related to the thermometer electron temperature (through equation 11) and is omitted on the plots for clarity. When the amplifier is powered (figure 5a), it first goes to saturation at $V_{out}=+3$ Volt. The electron temperature quickly increases, followed by the phonon bath. After few milliseconds, the electron and phonon temperatures reach the equilibrium value, 428.6 milliKelvin. The preamplifier output voltage then relaxes to its equilibrium value: 0.173 V. Figure 5b shows system reaction to a physical event. At time .5 milliseconds, the simulator puts a 10 keV heat pulse in the phonon bath. This warms up the electron's bath through electron-phonon coupling, and the preamplifier output drops until the electron temperature (and thermometer resistance) is restored to its equilibrium value. The simulated risetime fits the order of magnitude calculation of equation 10 and takes less than 10 µs. The pulse time duration is dominated by the thermal transfer time from the phonon to electron bath. The phonon's temperature relaxes through the electron-phonon thermal impedance with expected time constant: $C_{ph}/g_{eph}=4$ µs. Recovery of the equilibrium state takes overall less than 30 µs. The value of the offset voltage does not qualitatively change the behaviour of the system.

### 2. Stability

We notice that the Gain Bandwidth Product of the above simulated preamplifier is very high, close to the best performances achievable with discrete components. The reason is to prevent rigging between preamplifier output and electron temperature. As intuition tells us, to achieve fast smooth relaxation of the bolometer, the preamplifier's dominant pole has to be at higher frequency than the fastest relevant thermal frequency of the bolometer: $C_e/g_{eph}=7\ 10^{-7}$ s. We also want to have A/G≥100 to minimise errors. These requirements led us to choose such a high value for the Gain Bandwidth Product. This is a limitation of this configuration: some bolometers may turn out to be extremely difficult to readout. If $g_{eph}$ were 50 times larger, the bolometer would work in the adiabatic regime, and a pole preamplifier of lower frequency would do the job very well.

These simulations show that such a preamplifier is a valid configuration to readout a bolometer. Furthermore this system is a very fast temperature regulator, provided we can achieve good thermal coupling between the thermometer and the object to be regulated.

## D. Discussion

Table 1 summarises the computations.

The most commonly used bolometer readout configuration so far is the unlooped voltage amplifier. Its drawbacks are well known. Due to the input stray capacitance and high thermometer impedance, the signal is electronically low-passed. This results in the loss of the rise time information. Transient effects [3] in bolometers can hardly be studied. When used in the lock-in configuration, the parasitic capacitor limits the modulation frequency range, and induces a phase shift.

The looped voltage amplifier configuration moderately shortens the rise time. The limitation on it is technological. Most resistive thermometers at low temperature are based on the metal to insulator transition. Close to the transition, their dielectric constant increases. It turns out that unless thin film technology is available [2], it is very difficult to achieve thermometers with stray capacitance much lower than of 20 pF. Thus, we expect a rise time shortened by a factor 5, for the looped voltage configuration, with respect to unlooped configuration.

The looped current amplifier readout configuration solves the problem of rise time. Low temperature resistors of stray capacitance 0.2 pF are available on the market. Using 10 MΩ resistor, we expect a rise time of 4.4 μs, independent of the thermometer impedance. Much shorter transient effects can then be studied. When used in a lock-in amplifier loop, with square wave AC bias, the preamplifier provides a detected square wave with minimal distortion, (no amplitude loss, and a minute phase shift (delay) not dependent on the sensor impedance). Compensation of the phase shift may be unnecessary. If implemented, the adjustment is fixed, no tuning is necessary when sensor impedance changes. In case of spurious noise, the modulation frequency can be chosen in a much wider frequency range, without requiring a new calibration of the apparatus.

In practice, the looped current amplifier allows much flexibility of the design. The looped voltage amplifier configuration does not allow the best rise times, and since the preamplifier has a voltage gain of 1, requires a low noise second stage post amplifier to avoid performance's degradation of the electronic chain. This problem can be overcome by adding a gain divider of 1/C at the output of the operational amplifier and then closing the loop on the bolometer. The preamplifier output is then before the divider and the voltage gain becomes C. The drawback is that the dynamic range is then reduced.

It is to be pointed out that electrothermal relaxation is slower using the looped current amplifier configuration than the looped voltage amplifier. For experiments aiming at measuring high rate particle interactions, this is a drawback of the looped current amplifier configuration.

The impedance-regulating amplifier combines the advantages of fast rise time and shortens relaxation time. The advantage of fast rise time has been discussed at the previous paragraph. The new feature is that due to extreme electrothermal feedback the electrothermal relaxation time is dramatically reduced. This allows the use of conventional bolometers, to handle particle detection rates approching those of semiconductor diodes. Saturation of the output can be troublesome to measure the event energy through analog filtering. Fast sampling of signal should allow recovering information when this happens.

Furthermore, on time scales larger than a few $\tau_{eth}$, the output of the impedance regulating preamplifier directly measures the variation of the signal power entering the bolometer. The properties of the bolometer and preamplifier do not enter in equation 6. This means that the bolometer with its electronic readout chain is *auto calibrated*.

Finally, if $V_{eq}$ is a steady state, then $-V_{eq}$ is an other. If we achieve a design to switch from one state to the other, under the control of a clock, the thermometer becomes square wave AC biased. We then achieve the modulation and preamplifier stages of an auto biased lock-in detection chain. The preamplifier itself replaces the commonly used high-stability biasing generator. This design requires fewer electronic parts and does not require a very high stability generator ($10^{-6}$ relative stability in amplitude on a time scale of 1 second in previous design configurations).

The main drawback of the impedance-regulating amplifier is that the design requires understanding of the thermal behaviour of the bolometer to be reliable (choosing the bias point, main pole of the preamplifier and prevent ringing).

If noise is not a matter of concern, we can now choose the best configuration for the planned measurement, given the bolometer available. Bolometers however are mostly used because of their high sensitivity, and low noise. Then a discussion of the noise is necessary.

## V. Noise modelling and computations

In the following we will assume the preamplifier is an operational amplifier with infinite bandwidth and an open loop gain A. For clarity we also assume that there is no parasitic capacitor in parallel with the thermometer and the feedback resistor. As a support for discussion, we use a bolometer running at a temperature of 300 mK, with parameters extracted from [13].

In bolometric detection the signal measured is the incoming power on the bolometer. It is usual practice to quantify noise contribution in terms of Noise Equivalent Power (NEP). In the following we will compute the noise contributions of the most fundamental noises sources:

• The well known Johnson noise of the biasing resistors and thermometer

•The thermal noise of the bolometer, originating from thermal fluctuations of the absorber, weakly linked to cold bath by the heat leak of thermal conductivity g.

• Noise of the cold JFET: the current noise of a good cold (150K) JFET is very low. We will use the typical value of $10^{-16} A/Hz^{1/2}$. The voltage noise $e_n$ has been measured and will be taken for numerical computations to the value of 1.5 nV/Hz$^{1/2}$. [14,7]

• In looped readout configurations, the output voltage noise due to the front end JFET noise heats the bolometer [15]. We name these contributions the electrothermal noise power, respectively due to JFET's input voltage noise $P_{en}$ and input current noise $P_{in}$.

Mather shows [15], taking into account the power dissipation in the thermometer induced by the Johnson noise voltage and the bias current, the Johnson noise induces an NEP spectral density at the input of the bolometer given by:

$$\frac{dP_{jon}^2}{df} = \left| 4kTP_e(1+\omega^2\tau_{th}^2) \right| \left| \frac{g^2T^2}{\alpha^2 P_e^2} \right| = \left| \frac{4kg^2T^{5/2}}{\alpha^2 P_e^{3/2}}(1+\omega^2\tau_{th}^2) \right|$$

(W$^2$/Hz), where k is the Boltzmann constant, and T the temperature.

Mather computed also the thermal noise NEP spectral density:

$$\frac{dP_{therm}^2}{df} = \left| 4kT^2 g \right|$$

(W$^2$/Hz). For an impedance regulating amplifier, electronic noise contributions are calculated through the electrothermal: a noise voltage at the input of the amplifier produces a Joule power variation on the thermometer. This happens until the thermometer resistance (temperature) has changed to cancel the effect of the voltage noise.

**1. Electronic noise contributions.**

Let us define $\varepsilon_0$ the offset voltage at the input of the operational amplifier of figure 2d). Using conservation of charge, eliminating the current flowing through the capacitor $C_s$ and then I and making a first order expansion in $\varepsilon_0/V_{out}$ we get:

$$\frac{dR_b}{d\varepsilon_0} = \frac{R_b(G+1)}{V_{eq}}\left(1 + j\omega R_b\left(\frac{C_s G}{G+1} + C_b\right)\right).$$

Replacing $\tau_{eth}$ by its value in the definition of $H_{therm}$ we can write:

$$H_{therm}(j\omega) = \frac{R_b(G+1)}{2AP_e}\frac{1}{1+j\omega\tau_{eth}}.$$

Then the power necessary to fake $dR_b$ is:

$$dP_e = \frac{2AP_e}{R_0(G+1)}\left[1 + j\omega\tau_{eth}\right]dR_b$$

We get the NEP for the FET voltage noise:

$$\frac{dP_{en}^2}{df} = \left| \frac{2AP_e}{V_{eq}} (1 + j\omega\tau_{eth}) \left(1 + j\omega R_b \left(\frac{GC_s}{G+1} + C_b\right)\right) \right|^2 \frac{de_n^2}{df} \qquad (W^2/Hz)$$

In a similar way, a current noise establishes an offset voltage at the input of the operational amplifier. In the limit of large A:

$$\frac{d\varepsilon}{di} = \frac{R_b}{\frac{1+G}{G}\left(1 + j\omega R_b \left(\frac{GC_s}{G+1} + C_b\right)\right)}$$

From this we deduce the current noise contribution

$$\frac{dP_{in}^2}{df} = \left| \frac{2AP_e R_b G}{(G+1)V_{eq}} (1 + j\omega\tau_{eth}) \right|^2 \frac{di_n^2}{df} \qquad (W^2/Hz)$$

## 2. Discussion

For the previously introduced configurations, NEP spectral densities are classically extracted from references [15, 16]. Figure 6 displays the noise contributions as NEP spectral density at the input of the bolometer for the four configurations. For clarity, only the dominant noise contributions are plotted.

As expected, the thermal noise spectral densities are the same for the four readout configurations. At frequencies within the bandwidth of the bolometer, thermal noise dominates JFET noise contribution. This shows that JFETs have the potential to read bolometers at lower temperatures, with improved performances.

In the calculations the bolometers have been rather strongly biased. This allows us to demonstrate the impact of electrothermal feedback on signal readout. Figure 6 shows that the current amplifier depresses electronic noise contributions: $P_{in}$ becomes negligible (not shown) and $P_{en}$ is lower by a factor 10 relative to Voltage amplifier configurations. The reason is that the positive thermal feedback enhances the thermal noise $P_{therm}$, the Johnson noise relative to electronics contributions. The unlooped and looped voltage amplifiers display qualitatively the same behaviour, but the negative electrothermal feedback effect depresses the thermal signal amplitudes relative to electronic contribution.

The impedance-regulating amplifier increases the thermal bandwidth, but enhances the JFETs noise contributions. $P_{en}$ dominates all other noise contributions, degrading the resolution of the measurement. This is the same effect as using voltage amplifiers. In the impedance regulating configuration the negative electrothermal feedback is extreme. To prevent resolution loss, we wish to reduce $P_{en}$ noise spectral density at the level of the thermal noise. This requires optimisation of all the detection parameters. A/(G+1) has to be reduced to a value of the order of 100 and the bolometer impedance has to be matched to the cold JFET noise resistance, of the order of GigaOhm. This would slow down the preamplifier reaction time. Unless we use front end transistors (devices) with noise temperatures significantly reduced compared to today available JFETs, it seems unlikely we can get the full advantages of the impedance regulating amplifier with no loss in signal to noise ratio. This is can be a major drawback in low noise implementations.

## VI. Conclusion

We have studied the performances of four different readout electronic configurations for cryogenic bolometers. We have presented and analysed a new preamplifier configuration, the impedance regulating amplifier. Using impedance-regulating amplifiers, we achieve extreme electrothermal feedback electronically, using resistive thermometer and JFETs. Thus thermal relaxation times can be drastically reduced allowing particle detection rates comparable to semiconductor detectors. Bolometer and electronic detection chain are autocalibrated, the output voltage providing unambiguously the input power variation. The price to pay is that the noise contribution from the front-end transistor is enhanced. Experimental constraints will decide which looped readout configuration is the most convenient for a given apparatus and bolometer.

## VII. Acknowledgements

We want to thank the staff of the Service d'Electronique et d'Informatique (DAPNIA/SEI, Saclay) and Service d'Instrumentation Génerale (DAPNIA/SIG, Saclay) for their support. We are grateful to J.L Bret and A. Benoit, (CRTBT Grenoble), for stimulating discussions, to JP.Torre (Laboratoire d'aéronomie, Verrière le Buisson, France) and P. Garoche (Universite d'Orsay, France) for their help, to J. Rich (CEA/Saclay) for carefully correcting the manuscript proofs. We are happy to acknowledge the scientific input of Irwin et al.[17], that motivated us to develop the impedance-regulating amplifier. We want to thank D. Mc Cammon and the referies for fruitfull comments.

**References**

[1] E.E. Haller, N. P. Palaio, M.Rodder, W.L. Hansen, and E. Kreysa, in "Neutron Transmutation Doping of Semiconductor Materials", Plenum, New York, ed. R.D. Larrabée, pp. 21-36, 1984.

[2] S. Marnieros, L. Bergé, L. Dumoulin, A. Julliard and J. Lesueur, "Development of massive bolometers with thin film thermometers using ballistic phonons", Proc. of the VII International Workshop on Low Temperature Detectors (LTD-7), Max Planck Institute of Physics, Munich, pp. 134-136, July 1997.

[3] D. Yvon, L. Bergé, L. Dumoulin, P. De Marcillac, S. Marnieros, P. Pari, G. Chardin, "Evidence for signal enhancement due to ballistic phonon conversion in NbSi thin film bolometers", Nucl. Inst. Meth. in Phys. Res. A370, pp. 200-202, 1996.

[4] E. Silver, S. Labov, F. Goulding, N. Madden, D. Landis, J. Beeman, T. Pfafman, L. Melkonian, I. Millett, and Y. Wai, "High resolution x-ray spectroscopy using germanium microcalorimeters", "X-Ray, and Gamma-Ray Instrumentation for Astronomy and Atomic Physics", SPIE, vol. 1159 EUV, pp. 423-432, 1989.

[5] D. Drain and the EDELWEISS collaboration, "Status of the EDELWEISS experiment", Phys. Rep. 307, p. 297-300, 1998.

[6] ESA Planck surveyor project, updated information can be found on Web site: http://sci.esa.int/planck/

[7] D.V. Camin, G. Pessina, P.F. Manfredi., "Voltage-Sensitive Differential Input Preamplifier with Outstanding Noise Performances", Alta Frequenza 56, pp. 347-351, Oct 1987. and D. Yvon, A. Cummings, W. Stockwell, P.D. Barnes Jr., C. Stanton, B. Sadoulet, T. Shutt, C. Stubbs, "Low noise voltage and charge preamplifiers for phonon

and ionization detectors at very low temperature", Nucl. Inst. Meth. in Phys. Res. A 368, pp. 778-788, 1996.

[8] R. C. Jones, "General theory of bolometer performance", J. of the Opt. Soc. of Am. 43, pp. 1-14, 1953

[9] M. Galeazzi, F. Gatti, A. von Kienlin, P. Meunier, A.M. Swift, S. Vitale, "An Electronic Feedback System for improving the counting Rate of cryogenic Thermal Detector", Proc. of the "VII International Workshop on Low Temperature Detectors (LTD-7)", Max Planck Institute of Physics, Munich, pp. 174-175, July 1997.

[10] E. Aubourg, A. Cummings, T. Shutt, W. Stockwell, P.D. Barnes Jr et al., "Measurement of Electron-Phonon Decoupling Time in Neutron-Transmutation Doped Germanium at 20 mK", J. of Low Temp. Phys. 93, pp. 289-294, 1993.

[11] L. Dumoulin, CSNSM, IN2P3-CNRS, Bat 108, F-91405 Orsay Campus Cedex, France, private communication

[12] E. T. Swartz and R.D. Pohl, "Thermal boundary resistances", Rev. of Mod. Phys. 61, pp. 605-625, 1989.

[13] T. Wilbanks, "A Novel Measurement of the Sunyaev-Zel'dovich Effect in Galaxy Clusters", Ph. D. Thesis, University of California at Berkeley, (1994). Available at: http://astro.caltech.edu/~lgg/suzie/srefs.html.

[14] InterFET corp., 322 Gold Street, Garland Texas 75042, USA.

[15] J. C. Mather, "Bolometer noise: non equilibry theory", Appl. Opt. 21, pp. 1125-1129, 1982. J.C. Mather, "Bolometers: ultimate sensitivity, optimization, and amplifier coupling", Appl. Opt. 23, pp. 584-588, 1984.

[16] V. Radeka, "Low-noise techniques in detectors", Ann. Rev. Nucl. Part. Sci. 38, pp. 217-277, 1988.

[17] K.D. Irwin, G.C. Hilton, J.M. Martinis, B. Cabrera, "A hot-electron microcalorimeter for X-ray detection using a superconducting transition edge sensor with electrothermal feedback", Nucl. Inst. Meth. in Phys. Res. A 370, pp. 177-179, 1996.

**Captions:**

Table 1): Summary of the numerical values presented in the previous paragraphs, for the four readout configurations presented in this paper, assuming the detector parameters extracted from [8]. Within the adiabatic thermal model presented here, the electronics risetime would be the risetime of a pulse induced by a particle interaction in the bolometer, and $\tau_{eth}$, the relaxation time. The so-called thermal bandwidth is the effective bandwidth at -3dB, when measuring a varying physics power $\Delta P_{ph}(s)$, as deduced from $H_{tot}(s)$.

Figure 1): The adiabatic bolometer model. The detector is made from a thermometer in perfect thermal contact with an absorber at a temperature T. The total heat capacity is $C_t$, The absorber is thermalised to the cryostat through a heat leak of dynamic thermal conductance g=dP/dT.

Figure 2): Basic schematics of the four preamplifiers configurations we discussed in this paper.

Figure 3): Realistic thermal model of a bolometer, including electron-phonon decoupling in the thermometer [3, 5]. The thermal behaviour of the bolometer is modelled by the two following differential equations:

$$C_e \frac{dT_e}{dt} = P_{elec} - g_e V_s (T_e^{5.7} - T_{ph}^{5.7}) \text{ and } C_{ph} \frac{dT_{ph}}{dt} = g_e V_s (T_e^{5.7} - T_{ph}^{5.7}) - g_K (T_{ph}^4 - T_0^4) + P_{Phys}$$

where $T_e$ and $T_{ph}$ are the thermometer and electron temperatures, $T_0$ is the cryostat temperature, and $g_e$ and $g_0$ are the heat conductivities defined in text. The thermal impedance labelled "Kapitza" [12] is assumed to be negligible.

Figure 4): Pspice model used to simulate the impedance-regulating amplifier. The two first schematics simulated the thermal behaviour of the bolometer using the differential equations written at figure 3, but generalised to the non-linear case as discussed in the text. The third schematic is a model of a single pole operational amplifier, with input offset $V_{os}$, open-loop gain $A_0$ and cutting frequency $f_t$. $R_0$, delta and pr parameters fit the thermometer resistance versus temperature. $V_s$, ge, and pe, parameterise the electron-phonon thermal coupling. gk, pk parameterise the Kapitza resistance assumed to dominate the heat leak. $c_e$ and $c_{ph}$ parameterise the heat capacity of the electron and thermal bath. Part =0 means that the signal energy is deposited in phonon bath only. See text for further details.

Figure 5): Evolution with time of the computed output voltage, and the electron and phonon bath temperatures. Preamplifier output voltage is drawn as circles. Squares and diamonds represent respectively the electron and phonon temperature evolution. Figure 5a shows time evolution when the preamplifier is powered. Figure 5b shows preamplifier's response to a particle interaction in the bolometer's absorber.

Figure 6): Dominant noise contributions versus frequency, for the four readout configurations studied in this paper: a) unlooped voltage amplifier readout, b) the looped voltage amplifier, c) the looped current amplifier, d) the impedance regulating amplifier. Noise contributions are expressed as power noise equivalents at the input of the bolometer. $P_{en}$ is the noise due to the JFET voltage noise, $P_{in}$ is the noise due to JFET's current noise, $P_{John}$ is the noise contribution due to the thermometer's Johnson noise. $P_{therm}$ is the bolometer thermal noise.

| Readout configuration | Electronic risetime (s) | $\tau_{eth}$ (s) | Thermal Bandwidth (Rad s$^{-1}$) |
|---|---|---|---|
| Unloop Voltage amplifier | $2.2\ 10^{-3}$ | $3.3\ 10^{-3}$ | 280 |
| Looped Voltage amplifier | $4.4\ 10^{-4}$ | $3.3\ 10^{-3}$ | 300 |
| Looped current amplifier | $4.4\ 10^{-6}$ | $3.2\ 10^{-2}$ | 31 |
| Impedance regulating amplifier | $10^{-6}$ | $8\ 10^{-7}$ | $1.25\ 10^{6}$ |

Table 1)

| PARAMETERS: | | PARAMETERS: | | PARAMETERS: | | PARAMETERS: | | PARAMETERS: | | PARAMETERS: | | PARAMETERS: | |
|---|---|---|---|---|---|---|---|---|---|---|---|---|---|
| R0 | 1900 | Vs | 2.5E-7 | gk | 1e-8 | ce | 6E-5 | Teq | 428E-3 | fac | 10 | E0 | 1.6E-16 |
| delta | 7.72 | ge | 280 | pk | 4 | cp | 1E-9 | Teq | 428E-3 | delay | 0.5ms | part | 0 |
| pr | 0.64 | pe | 5.7 | | | | | T0 | 280E-3 | feedb | 90 | | |

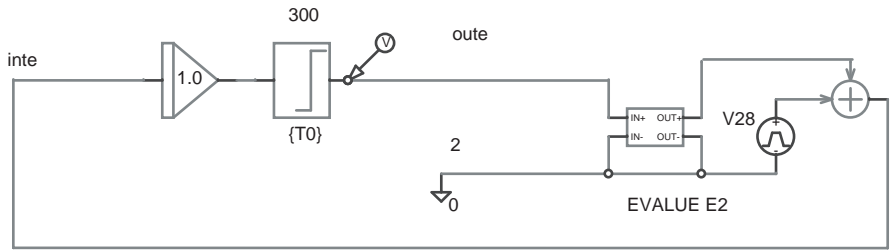
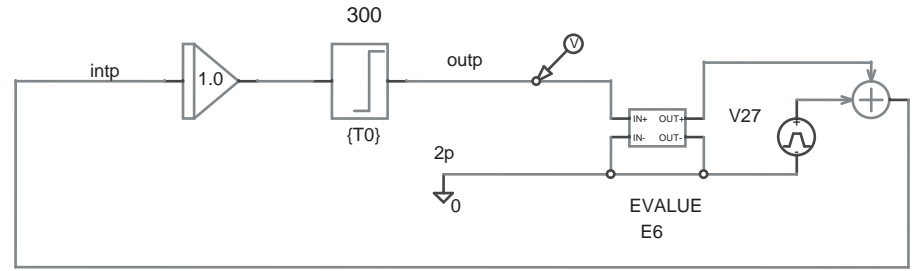

(ge*Vs/(cp*pwr(v(outp,2p),3)))*(pwr(v(oute,2),pe)-pwr(v(outp,2p),pe))-(gk/(cp*pwr(v(outp,2p),3)))*(pwr(v(outp,2p),pk)-pwr(T0,pk))

pwr(v(a,b),2)/(R0*exp(pwr((delta/v(oute,2)),pr))*(v(oute,2)*ce*Vs))-(ge/(ce*v(oute,2)))*(pwr(v(oute,2),pe)-pwr(v(outp,2p),pe))

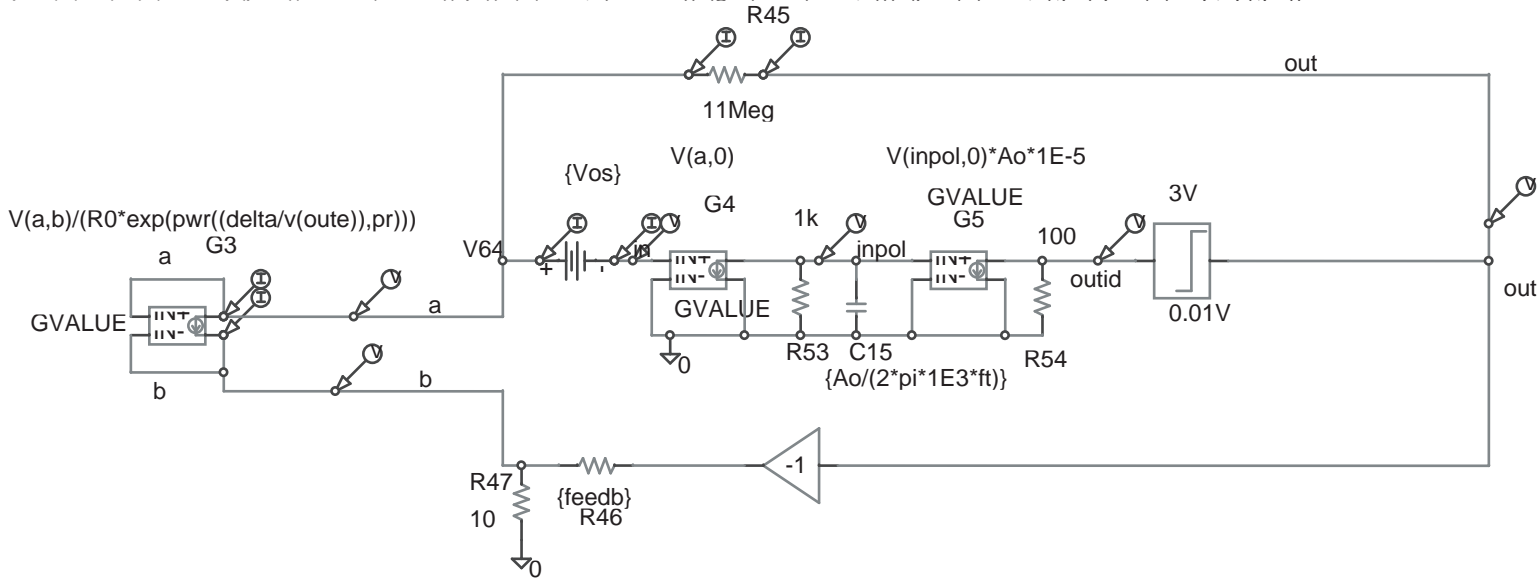

PARAMETERS:
Vos  0.1mV
Ao   0.01Meg
ft   2000Meg

PARAMETERS:
pi  3.141592654

Figure 4)

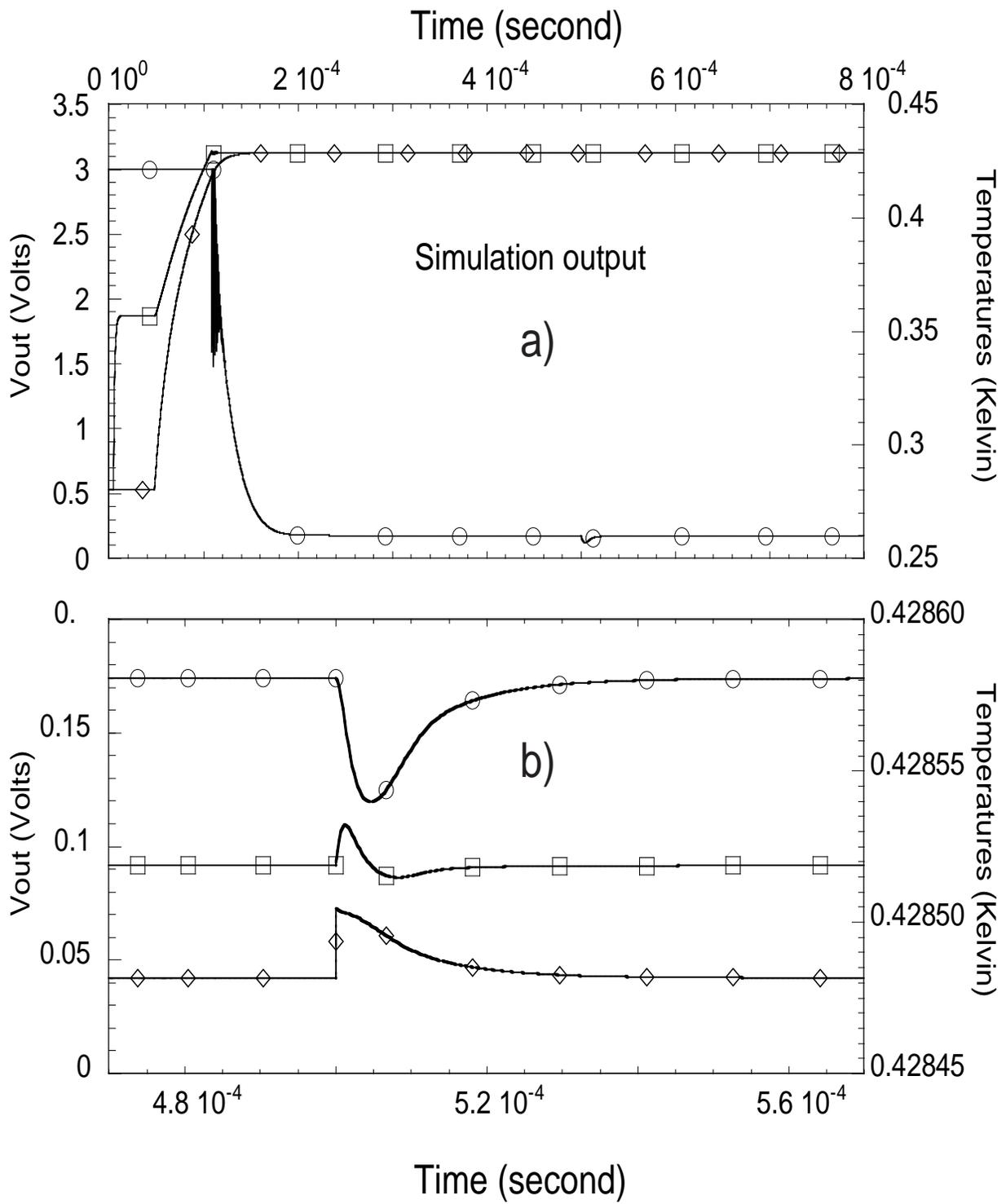

Figure 5)

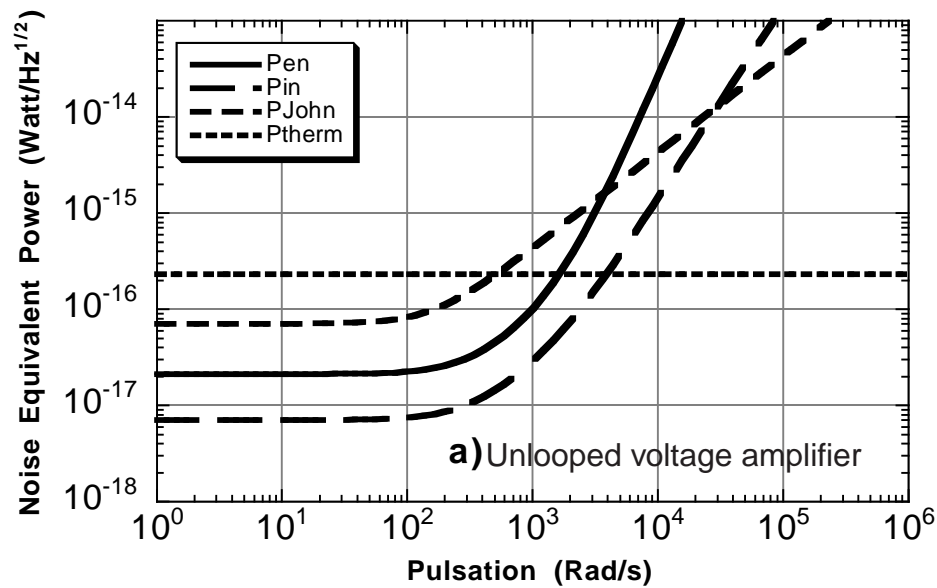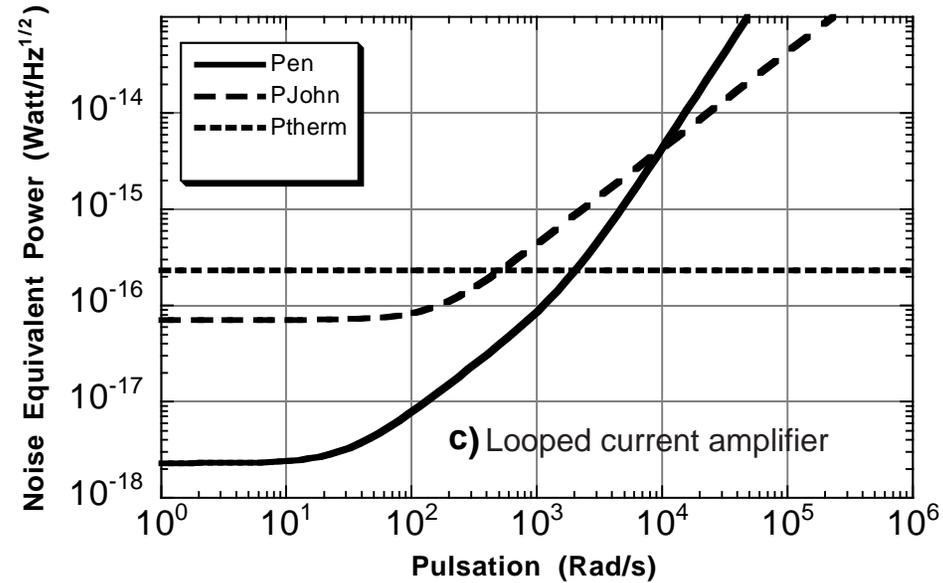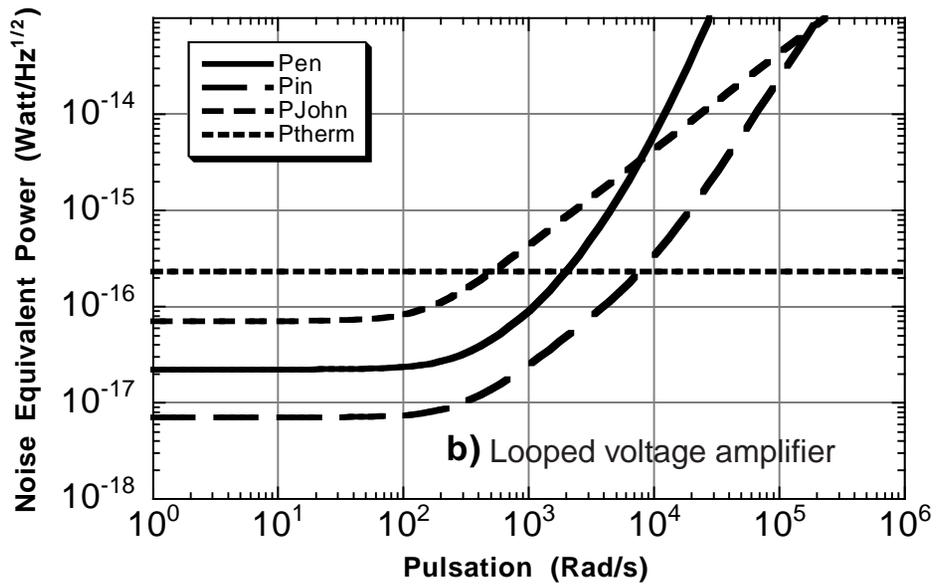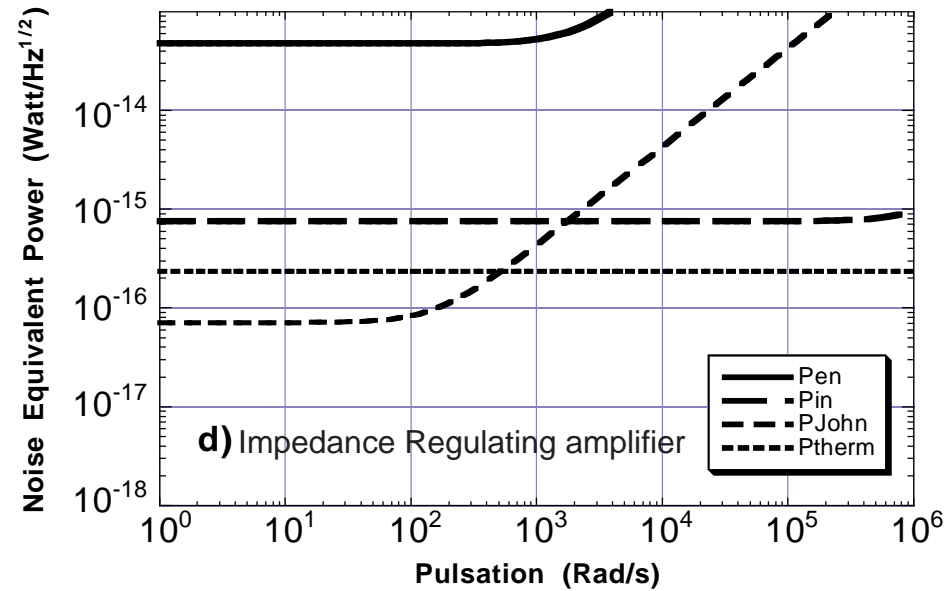

Figure 6)